# Autonomous Agent-Orchestrated Digital Twins (AADT): Leveraging the OpenClaw Framework for State Synchronization in Rare Genetic Disorders

Hongzhuo Chen[1,2], Zhanliang Wang[1,2], Quan M. Nguyen[1,3], Gongbo Zhang[4], Chunhua Weng[4], Kai Wang[1,5*]

[1] Raymond G. Perelman Center for Cellular and Molecular Therapeutics, Children's Hospital of Philadelphia, Philadelphia, PA 19104, USA

[2] Applied Mathematics and Computational Science Graduate Program, University of Pennsylvania, Philadelphia, PA, 19104, USA

[3] Bioengineering Graduate Program, University of Pennsylvania, Philadelphia, PA 19104, USA

[4] Department of Biomedical Informatics, Columbia University Irving Medical Center, New York, NY 10032, USA

[5] Department of Pathology and Laboratory Medicine, Perelman School of Medicine, University of Pennsylvania, Philadelphia, PA 19104, USA

*: Correspondence should be addressed to wangk@chop.edu.

# Abstract

**Background:** Medical Digital Twins (MDTs) are dynamic computational representations of individual patients, integrating clinical, genomic, and physiological data to support diagnosis, treatment planning, and outcome prediction. MDTs allow physicians to use a virtual copy of a patient to improve disease diagnosis, predict disease trajectory, simulate drug response, and offer personalized clinical decisions. However, a significant synchronization gap persists: most MDTs are static or only updated passively. This represents a particular challenge in rare genetic disorders, where patient phenotypes, genomic data reinterpretation, and care guidelines are all evolving over time.

**Methods:** We propose an agent-orchestrated digital twin framework utilizing OpenClaw's proactive "heartbeat" mechanism and modular Agent Skills. This Autonomous Agent-orchestrated Digital Twins (AADT) framework is engineered to autonomously monitor local and global data streams (e.g., self-reported phenotype updates and evolving variant classification in public databases), execute automated bioinformatics workflows to manage data ingestion, normalization, state updates, and workflow triggers.

**Results:** Prototype implementation suggests that an agent-orchestrated approach can be used within OpenClaw to continuously synchronize the digital twin's state with both patient-reported phenotypic observations and external genomic knowledge updates. In the context of rare genetic disorders, this proactive synchronization may result in earlier genetic diagnosis of diseases and more accurate modeling of the natural history of diseases. We use two case studies on variant reinterpretation and longitudinal phenotypes to show how AADTs facilitate research and clinical care of rare genetic diseases, through timely and auditable updates.

**Conclusion:** The AADT framework addresses the critical bottleneck of real-time data collection, potentially allowing more scalable, data-rich establishment of MDTs. We further discuss challenges in data security and mitigation strategies with the human-in-the-loop design.

# Introduction

Rare genetic disorders collectively affect hundreds of millions of individuals worldwide[1-3], yet each condition individually faces challenges stemming from low prevalence, heterogeneous phenotypic trajectories, and limited clinical expertise. Patients frequently undergo prolonged diagnostic odysseys, iterative reassessments, and evolving management strategies as new genomic evidence, phenotypic data, and therapeutic options emerge[4-7]. Care is often fragmented across institutions and specialties, complicating longitudinal understanding and coordination. These characteristics make rare diseases a compelling but demanding domain for emerging digital health paradigms.

The concept of a medical digital twin (MDT)[8,9], which is a computational representation of an individual that evolves over time in response to new data, has gained increasing attention as a potential mechanism for addressing such complexity. Across healthcare domains, MDT has been proposed for physiological simulation, treatment response prediction, and personalized decision support[10-15]. Recent reviews emphasize their promise for enabling precision medicine by integrating multi-scale data, including clinical measurements, imaging, and molecular profiles[16-19]. However, these same reviews consistently highlight unresolved challenges related to data integration, validation, governance, and real-time updating[20-22]. In practice, many MDT implementations remain static, retrospective, or narrowly scoped to specific modeling tasks[23].

The application of MDT to rare disease has been discussed extensively in literature[24-28]. We believe that rare diseases are particularly well suited to digital twin approaches due to their longitudinal nature (clinical phenotypes change over time), evolving diagnostic interpretations, and reliance on heterogeneous data sources (imaging, EEG, genetics, transcriptomics, etc.)[24,25]. At the same time, there are known obstacles: how to maintain a continuously updated representation of a patient; how to reconcile conflicting or uncertain genetic evidence; and how to integrate such systems into real research protocols or clinical workflows[24]. Therefore, despite broad agreement on the potential value of MDT for rare diseases, concrete implementation frameworks addressing these challenges remain limited[24,27].

Parallel to developments in MDT research, recent advances in personal AI assistants and agent-based systems have introduced new capabilities for orchestrating complex workflows across digital environments[29,30]. These systems are designed to interact with users, invoke tools, monitor events, and execute tasks across heterogeneous data sources. In non-clinical settings, agent runtimes have demonstrated utility for automation, scheduling, and information synthesis. Their flexibility and event-driven nature suggest a potential role as orchestration layers for continuously updating digital representations, including medical digital twins.

Building on this premise, we adopt OpenClaw as the agent runtime for our framework. OpenClaw is an open-source autonomous AI agent framework designed to function as a "local-first" personal assistant that resides directly on a user's hardware. Unlike conventional prompt-driven AI assistant, OpenClaw operates with a proactive heartbeat mechanism that allows it to independently moniter local file systems and browse the web on a predefined schedule after initial configuration. For researchers and clinicians, OpenClaw acts as an autonomous orchestrator by bridging the gap between raw data and actionable insights; it can be programmed via modular "Agent Skills" to watch for new database or literature information, trigger local bioinformatics pipelines, and automatically synchronize these findings with a MDT. By maintaining a long-term, version-controlled memory in local Markdown files, it ensures that clinicians receive real-time updates on a patient's disease trajectory while maintaining strict data sovereignty and privacy.

In this work, we propose an autonomous agent-orchestrated digital twin (AADT) framework tailored to rare genetic disorders, leveraging OpenClaw as the backbone. Our goal is not to introduce a new disease-specific MDT model, but rather to develop a practical, standards-aware architecture for the informatics infrastructure required to keep an MDT current, traceable, and usable. We define an *Autonomous Agent-orchestrated Digital Twin* (AADT) as an event-driven architecture in which a constrained personal AI assistant runtime serves as an orchestration layer. We further outline two common use cases to illustrate how this strategy can improve current workflow of variant reinterpretation and continuous clinical evaluation for rare undiagnosed diseases.

## Results

### System Feasibility and Architecture

The core contribution of this work is an architecture that shifts the Medical Digital Twin (MDT) from a static record to an active, synchronized model. This framework is orchestrated by the OpenClaw agent, which serves as the intelligent middleware between the patient's physical data and its virtual representation. An illustration of the AADT framework is given in Figure 1. Briefly, the OpenClaw agent operates as a persistent local daemon. Its primary role is to bridge the "synchronization gap" through two mechanisms: (1) skill-based execution, through which the agent invokes specialized bioinformatics routines rather than relying on generalized LLM outputs; and (2) a proactive heartbeat, a scheduled polling cycle that monitors local data directories or remote databases for changes, triggering relevant skills for downstream analysis when updates are detected. For the purpose of this study, we implemented a single OpenClaw skill, PhenoSkill, which encapsulates two complementary functions: a phenotype monitoring task that detects emerging phenotypes in caregiver-submitted text and saves them as timestamped PhenoPacket records, and a variant reinterpretation task that autonomously queries the latest ClinVar releases.

A key design consideration is ensuring reproducibility, traceability, and auditability of the information generated by the agent. To achieve this, all data extraction is performed by PhenoSnap, a local, deterministic command-line toolkit, rather than by the large language model that powers OpenClaw. Therefore, regardless of the underlying LLM (such as GPT, Kimi or MiniMax) used by OpenClaw, the results remain identical. This concept is illustrated in Figure 1: while OpenClaw's underlying model handles skill activation and user communication, the actual clinical data processing, including phenotype extraction and variant reclassification lookup, is performed exclusively by PhenoSnap's deterministic Python scripts. We evaluate this architecture through two case studies and a minimal viable digital twin implementation. Case Study 1 demonstrates how longitudinal phenotype tracking through caregiver-reported observations can progressively refine a diagnostic ranking. Case Study 2 illustrates how automated monitoring of ClinVar reclassifications can transform a previously inconclusive genetic result into a confirmed molecular diagnosis. Both case studies converge on a minimal viable rare-disease MDT (RDMDT), which integrates the updated components into a unified digital twin state.

**Case Study 1: Emergence of new clinical phenotypes for rare undiagnosed diseases**

This case study shows a patient whose clinical presentation evolved gradually over the first few years of life, with initial symptoms too nonspecific to support a definitive diagnosis. The patient is a newborn male who presented with hypotonia and feeding difficulties at birth. These findings are common across a broad range of neurodevelopmental and genetic conditions. At this stage, the correct diagnosis (Angelman syndrome, caused by loss of function of the maternally inherited UBE3A gene) ranked 327th in phenotype-based disease prediction, offering limited diagnostic specificity.

The family enrolled in the AADT framework shortly after birth (Figure 2). Over the following months, the caregiver submitted periodic free-text updates describing newly observed symptoms through the OpenClaw interface. Each submission was automatically processed by PhenoSkill, which extracted standardized HPO terms from the caregiver's natural language descriptions and appended them to the patient's PhenoPacket record within the digital twin. At six months, the caregiver reported poor head control and global developmental delay. By twelve months, additional features emerged including strabismus, inability to sit independently, sleep disturbance, and excessive drooling, with the correct diagnosis rising to 91st in phenotype-based ranking. At eighteen months, the clinical picture became more distinctive: the caregiver described repetitive hand flapping, stereotypic behaviors, absence of speech, and seizure-like episodes, elevating the ranking to 30th.

At the age of two, the patient developed confirmed seizures, ataxia, and abnormal EEG findings, elevating the correct diagnosis to 4th in phenotype-based ranking. Standard genetic testing at this stage returned negative results. By three years, the emergence of highly characteristic features, including a happy demeanor with frequent inappropriate laughter, secondary microcephaly, and severe intellectual disability, brought the ranking to 2nd. A subsequent methylation test confirmed the diagnosis of Angelman syndrome. Notably, several features that appeared after diagnosis, such as wide mouth, prognathism, and scoliosis at four years, were retrospectively consistent with the established diagnosis and were captured in the digital twin for ongoing clinical management, at which point the correct diagnosis reached 1st in ranking. Throughout this process, the AADT framework maintained a complete, timestamped phenotypic trajectory, enabling clinicians to review not only the final HPO profile but the full chronological sequence in which each feature was first observed and reported.

**Case Study 2: Genomic Reinterpretation Enabling a Positive Diagnosis of a Rare Disease**

In this illustrative case study, we examined a patient with an inconclusive genetic testing result due to the presence of a variant of uncertain significance (VUS). Given the patient's clinical presentation, the clinical team ordered comprehensive genomic testing via whole-exome sequencing (WES). The sequencing report identified a variant of interest: NM_001142800.2(EYS):c.2137+1G>A, a splice-site variant reported in association with Retinitis pigmentosa (ClinVar accession RCV000779518.8), classified as VUS due to insufficient functional evidence, limited independent submissions in ClinVar, and low review confidence. The patient was counseled that periodic reinterpretation would be advisable. However, under conventional workflows, such reinterpretation typically relies on manual, clinician-initiated queries that could be difficult to sustain over time.

Upon enrollment in the AADT framework, a digital twin was created for the patient (Figure 3). The patient's VCF file containing the VUS was stored within the digital twin. OpenClaw's heartbeat mechanism then activated the ClinVar Monthly Scan skill, a scheduled agent task that periodically queries the ClinVar database for reclassifications of the patient's variants to Pathogenic or Likely Pathogenic at a minimum two-star review confidence.

For the first several months, each scan detected no reclassification events and the results were silently logged. Eight months later, the scan detected that the ClinVar classification of NM_001142800.2(EYS):c.2137+1G>A had been updated to Likely Pathogenic, following new functional splicing assay data and concordant submissions from multiple independent laboratories. The skill generated a structured alert containing the updated ClinVar significance, review status, and star rating, which was surfaced to the clinical team alongside the patient's stored phenotypic profile. This reclassification prompted a comprehensive genotype-phenotype concordance review, advancing the patient's diagnostic evaluation. All prior scan results and the evidentiary basis for the reclassification were preserved in the event log. With OpenClaw's heartbeat mechanism, which can deliver alerts directly to clinicians' and patients' mobile devices, this reclassification was surfaced at the earliest possible opportunity, facilitating timely clinical follow-up.

### Rare Disease Medical Digital Twin (RDMDT)

As part of the system implementation, we implemented a minimal viable rare-disease medical digital twin (RDMDT). The RDMDT integrates multimodal inputs including genotype data (VCF files with variant classifications), phenotype data (longitudinal HPO term profiles extracted by PhenoSnap), and auxiliary clinical data (e.g., EEG findings, imaging reports). The digital twin state is maintained as a single JSON file, assembled by reading all available data files from a designated patient directory. When any individual component is updated, for example, when PhenoSkill generates a new PhenoPacket file with additional HPO terms in Case 1, or when the ClinVar Monthly Scan detects a variant reclassification in Case 2, the corresponding file in the patient directory is overwritten and the twin state is reassembled to reflect the latest clinical picture. Components for which no data is available are left empty, allowing the digital twin to be instantiated even when only partial information exists.

## Materials and Methods

### OpenClaw Agent Runtime

OpenClaw (https://github.com/openclaw/openclaw) is an open-source autonomous AI agent framework designed as a local-first personal assistant that runs directly on user hardware. In this study, we used OpenClaw version 2026.3 (state release series), installed via the Node.js package manager from the official npm registry. OpenClaw provides two main capabilities to the AADT framework: (1) a proactive "heartbeat" mechanism, which is a scheduled polling cycle that allows the agent to autonomously wake up and monitor local file systems and remote databases without requiring human prompting; (2) a modular AgentSkill system, which enables domain-specific automation routines to be packaged, installed via ClawHub (the built-in skill registry), and registered within a workspace. The agent communicates with users through messaging interfaces (e.g., WhatsApp, Telegram, Slack) and maintains long-term, version-controlled memory in local Markdown files within a designated workspace directory. All agent operations in this study were executed locally, with no patient data transmitted to external servers.

## Human Phenotype Ontology and Phenotype Extraction

The Human Phenotype Ontology (HPO) provides a standardized vocabulary of phenotypic abnormalities observed in human disease. We used the HPO release version 2025-10-22, downloaded in OBO format (hp.obo) from the official OBO Foundry repository (http://purl.obolibrary.org/obo/hp.obo). The HPO file was stored locally within the OpenClaw workspace and parsed to extract all term labels and synonyms for use in phrase matching. PhenoSnap (https://github.com/WGLab/PhenoSnap) is a deterministic Python-based phenotype extraction tool developed as part of this study. PhenoSnap uses spaCy's PhraseMatcher to match clinical phrases in free-text input against the full set of HPO labels and synonyms parsed from the locally stored HPO ontology file. The pipeline additionally performs negation detection via dependency parsing to correctly handle excluded phenotypes (e.g., "no history of seizures" is marked as absent rather than present). All extracted phenotypes are output as timestamped PhenoPacket-compliant JSON records following the GA4GH Phenopacket schema, stored locally within the digital twin workspace.

## ClinVar Database and Variant Reinterpretation

ClinVar is a freely accessible public archive of reports describing the relationships among human variations and phenotypes, hosted by the National Center for Biotechnology Information (NCBI). ClinVar releases updated variant classifications on an approximately monthly cadence. In the AADT framework, we utilize PhenoSnap's variant lookup capability to perform automated variant monitoring against ClinVar through the NCBI E-utilities API. Specifically, for each variant stored in the patient's VCF file within the digital twin, the variant is first converted to SPDI (Sequence, Position, Deletion, Insertion) notation using RefSeq accessions following either the GRCh37 or GRCh38 reference assembly. The formatted variant identifiers are then submitted to the E-search endpoint (https://eutils.ncbi.nlm.nih.gov/entrez/eutils/esearch.fcgi) to retrieve matching ClinVar record IDs, followed by the E-summary endpoint (https://eutils.ncbi.nlm.nih.gov/entrez/eutils/esummary.fcgi) to obtain the current clinical significance classification and review status for each record. A reclassification event is defined as a change from Variant of Uncertain Significance (VUS) to either Pathogenic or Likely Pathogenic, with a minimum two-star review confidence, ensuring that only classifications supported by concordant submissions from multiple independent laboratories are retained.

## Agent Skill Implementation

A custom OpenClaw skill was developed for this study. PhenoSkill is an OpenClaw Agent Skill (installable from ClawHub; source code at https://github.com/kaichop/phenoskill) responsible for two complementary functions. First, when triggered by user input containing phenotype or medication cues (detected by OpenClaw's underlying language model), PhenoSkill invokes the PhenoSnap extraction program on the redacted input text and appends the resulting HPO terms with timestamps to the patient's longitudinal Phenopacket record. Importantly, prior to phenotype extraction, the skill scans the input for personally identifiable information and redacts obvious identifiers such as email addresses, phone numbers, and medical record numbers before writing the text to disk. Second, PhenoSkill implements a heartbeat-driven ClinVar Monthly Scan task, configured as a time-triggered execution cycle that runs automatically once per calendar month, aligned with ClinVar's monthly release cadence. During each cycle, PhenoSnap reads the patient's VCF file stored within the digital twin, converts each variant to SPDI notation, and queries ClinVar through the NCBI E-utilities as described above. If a reclassification event is detected, a structured alert is generated and surfaced to both the clinical team and the patient via the OpenClaw messaging interface.

## Discussion

In this work, we present the Autonomous Agent-orchestrated Digital Twin (AADT) framework, which is an informatics infrastructure that leverages an event-driven personal AI agent runtime to maintain continuously updated, auditable, and longitudinally coherent digital representations of patients with rare genetic disorders. Rather than introducing a disease-specific predictive model, our contribution focuses on the orchestration: we demonstrate that a constrained local AI assistant operating through scheduled proactive cycles and modular skill-based execution can systematically bridge the synchronization gap that makes most existing medical digital twin implementations clinically impractical. Two case studies illustrate this principle: autonomous genomic variant reinterpretation and longitudinal phenotype tracking in Angelman syndrome both demonstrating how agent-mediated infrastructure reduces the latency between the emergence of new clinical evidence and its incorporation into an actionable patient representation.

This synchronization gap reflects a deeper structural problem. Prior work has characterized this challenge primarily in terms of data heterogeneity and interoperability[31,32], but comparatively little attention has been devoted to the orchestration mechanisms required to sustain longitudinal synchronization in practice. We address this using an event-sourced state management architecture, in which each digital twin update is recorded as an immutable, timestamped event traceable to its originating source. This preserves the full provenance chain, allowing clinicians to reconstruct the evidentiary basis for any past recommendation and to identify precisely when and why a diagnostic interpretation changed. This property is especially important in genomic medicine, where variant pathogenicity classifications are revised with substantial frequency as knowledge bases such as ClinVar accumulate new evidence[33,34], and where current reinterpretation workflows are often reactive and incompletely communicated to treating clinicians.

However, trustworthy updates require careful delineation of what the agent is permitted to do autonomously and what must remain under human control. A central design principle of the AADT framework is constrained agency: the OpenClaw agent monitors, proposes, and escalates, but does not execute clinical decisions. This is consistent with emerging regulatory frameworks for AI-assisted clinical decision support, which hold that for consequential medical decisions, AI systems should augment rather than supplant human judgment [35,36]. The genomic reinterpretation case study instantiates this principle directly: the agent detected a variant reclassification event, generated a structured summary, and surfaced it for clinician review. The phenotype monitoring case study demonstrates a complementary value. In Angelman syndrome, the decisive diagnostic pivot depends on recognizing phenotypic crystallization over time rather than on sequencing alone[37,38], as standard microarray and whole-exome sequencing are unable to find the underlying imprinting defects (the Angelman syndrome can be caused by single nucleotide variants, structural changes or imprinting defects[38,39]). By continuously mapping patient-reported observations to HPO terms and maintaining an updated differential diagnosis, the framework reduced the interval between phenotype emergence and diagnostic re-prediction from a typical multi-week clinical review cycle to under 15 minutes, with meaningful implications for families navigating the rare disease diagnostic odyssey[40].

Constraining agent autonomy addresses one class of risk, but the deployment of autonomous agents in healthcare introduces additional security concerns that warrant explicit treatment. Extensible agent ecosystems are vulnerable to prompt injection, misconfigured permissions, and inadvertent exfiltration of protected health information[41,42]. Our local-first architecture and least-privilege execution model mitigate these risks by ensuring that patient data remains under institutional control and that each agent action is constrained to its minimum required permissions. A further concern is LLM hallucination[43]. We address this by restricting the LLM to orchestration tasks such as event detection, natural language understanding, and workflow coordination, while all clinical data extraction and transformation are performed by deterministic, version-controlled software tools. This separation of AI-mediated inference from structured data production provides a principled safeguard against hallucination-induced errors entering the digital twin state.

These design choices produce a system that is feasible and governable in prototype, but several limitations still exist. The results do not permit inference about real-world impact on patient outcomes or clinical efficiency, as the case studies are illustrative rather than prospective. The framework presupposes interoperable data sources, local computing infrastructure, and institutional governance mechanisms that may not be uniformly available. The current implementation also lacks formal mechanisms for federated or multi-institutional deployment, limiting opportunities for rare disease cohort aggregation. Prospective clinical validation across diverse care settings remains the most immediate priority. Longer term, integration with federated learning infrastructure and incorporation of validated predictive models including foundation models trained on population-scale EHR data[44,45], will require careful attention to uncertainty quantification and model governance.

The primary bottleneck to clinical digital twin adoption is not the absence of sophisticated predictive models[13,22]. It is the absence of reliable infrastructure for maintaining patient representations that are simultaneously current, auditable, and safe. The AADT framework provides a practical foundation for addressing this challenge, and the principles of longitudinal state preservation, provenance traceability, and human oversight articulated here will remain essential as the field advances.

# Acknowledgements

We thank the developers of the OpenClaw framework for providing the underlying agentic architecture that enabled the autonomous data management and digital twin modeling in this study. This study was supported by NIH/NHGRI grant HG013031 and the CHOP Research Institute.

# Author contributions

K.W. conceived the study. H.C. and K.W. developed the methodology and software. G.Z. and C.W. advised on the methodology development. H.C., Z.W. and Q.N. performed experiments and case studies. K.W. and H.C. wrote the original draft. Z.W., Q.N., G.Z. and C.W. reviewed and revised the manuscript. K.W. supervised the project.

## Declarations

### Competing interests

The authors declare no competing interests.

### Data and code availability

The phenoskill can be installed from ClawHub (source code available at https://github.com/WGLab/phenoskill). The set of phenotype characterization and genotype reinterpretation and digital twin state characterization tools can be found at https://github.com/WGLab/PhenoSnap. The HeartBeat file for genotype reinterpretation can also be found at https://github.com/WGLab/PhenoSnap and can be automatically installed by OpenClaw.

# Figures

**Figure 1. Overview of the Autonomous Agent-orchestrated Digital Twin (AADT) framework.** The OpenClaw agent provides two core mechanisms: skill-based execution for invoking specialized bioinformatics routines, and a proactive heartbeat for scheduled monitoring of local data directories and remote databases. PhenoSkill encapsulates two functions: phenotype monitoring, which extracts HPO terms from caregiver-submitted text, and variant reinterpretation, which queries ClinVar for reclassifications on a monthly cycle. All clinical data processing is performed by PhenoSnap, a deterministic command-line toolkit, ensuring reproducibility and auditability. Outputs are integrated into a unified rare-disease medical digital twin (RDMDT) state, evaluated through two case studies on longitudinal phenotype tracking and automated variant reclassification detection.

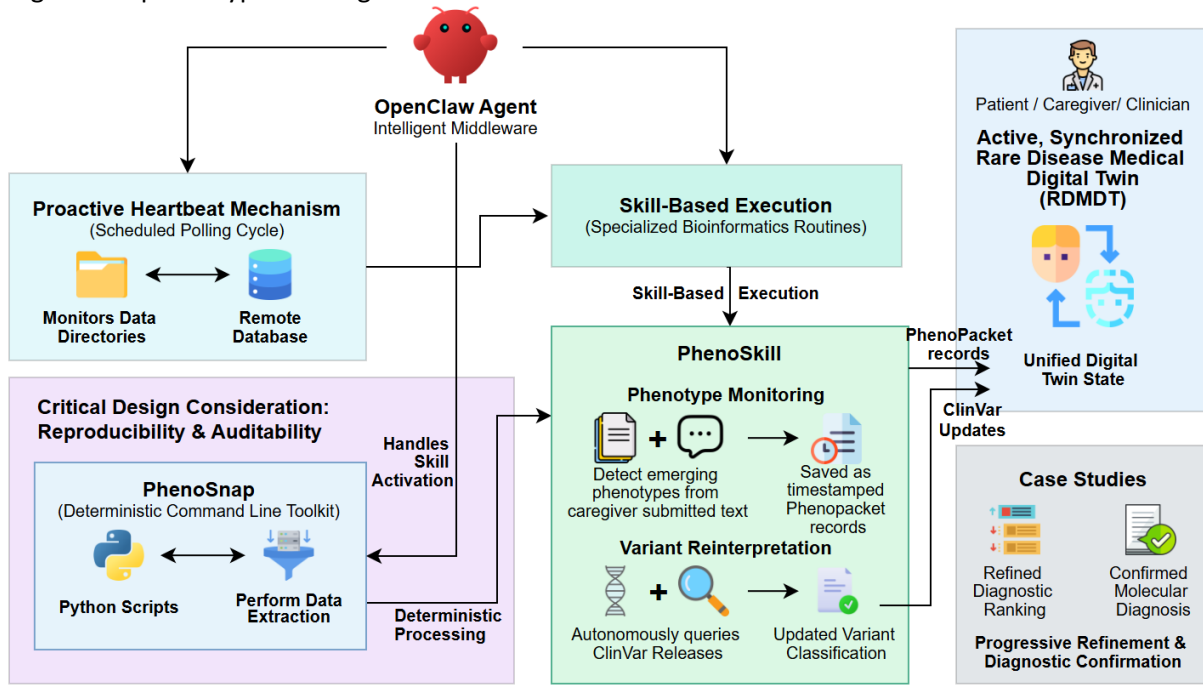

**Figure 2. Longitudinal phenotype tracking and progressive diagnostic refinement in Case Study 1.** Caregiver free-text symptom descriptions are submitted through the OpenClaw interface and processed by PhenoSnap, including PII redaction, HPO phrase matching, and negation detection, producing timestamped PhenoPacket JSON records stored within the digital twin workspace. The consolidated clinical timeline shows four representative stages of phenotypic evolution from birth through four years of age, with newly observed features extracted and appended to the patient's longitudinal HPO profile at each stage. As the phenotypic profile accumulates over time, the phenotype-based diagnostic ranking of the correct diagnosis (Angelman syndrome, UBE3A) progressively improves from 346th at 6-month old to 1st following post-diagnosis confirmation.

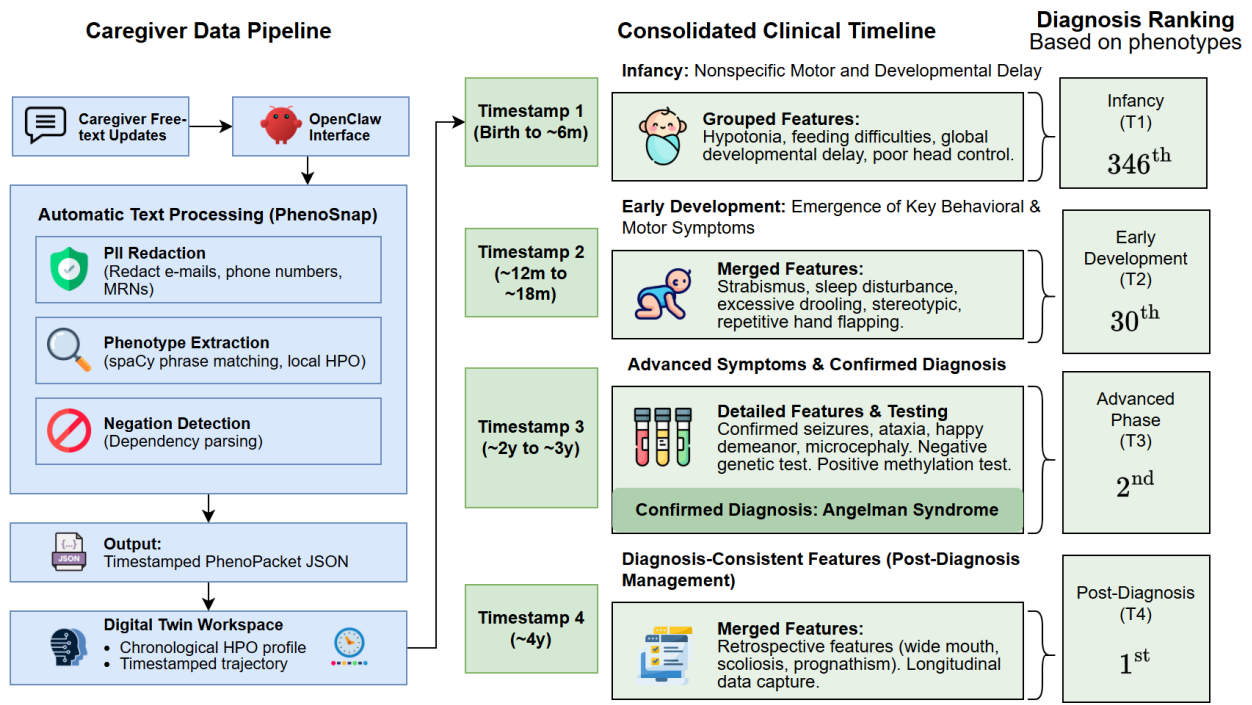

**Figure 3. Autonomous agent-driven variant reinterpretation and dynamic synchronization of the digital twin.** The patient's digital twin, maintained within the AADT framework, stores genomic and clinical data, including the VCF file, clinical profile, and reference assembly information. A heartbeat-driven scan executes automatically once per calendar month. During each cycle, variants are converted to SPDI notation using RefSeq accessions and queried against the ClinVar database through NCBI E-utilities: the E-search endpoint retrieves matching ClinVar record IDs, and the E-summary endpoint returns current clinical significance and review status. The skill filters for variants reclassified as Pathogenic or Likely Pathogenic, requiring a minimum two-star review confidence supported by concordant submissions from multiple independent laboratories. If no reclassification is detected, the result is silently logged to a persistent state file. When a reclassification event is identified, a structured alert containing the updated classification is surfaced to both the clinical team and the patient for timely reassessment.

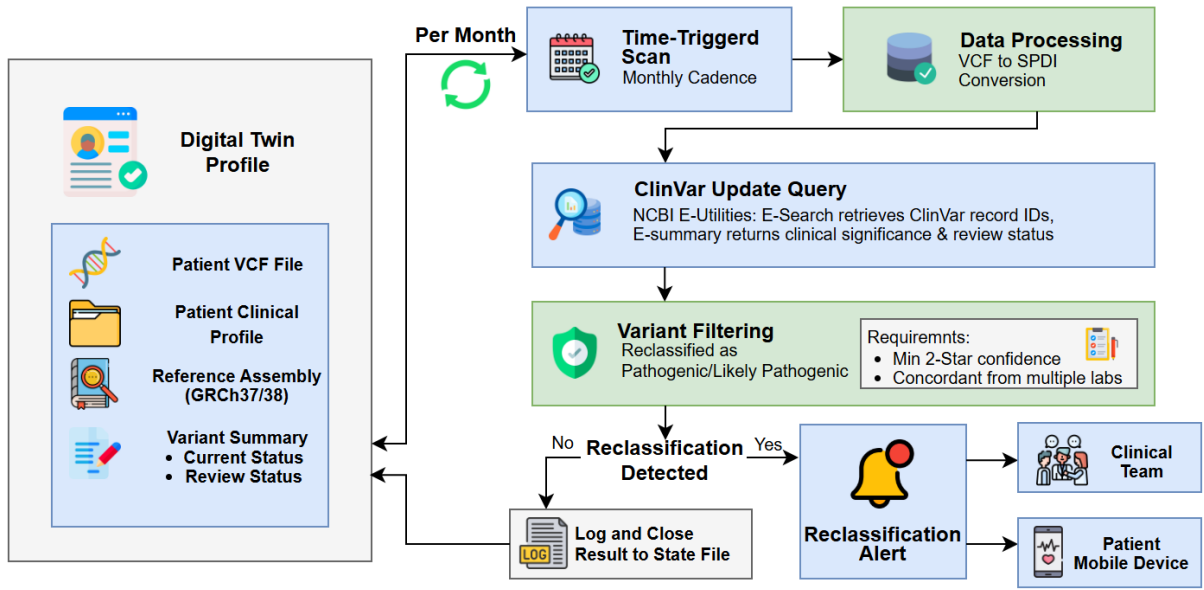

# Tables

Table 1. **Evolving phenotypes for a rare disease. The raw texts used to generate this table was given in Supplementary Notes.**

| Timestamp | HPO Terms | Top disease/gene prediction (UBE3A) |
|---|---|---|
| At Birth | <ul><li>Hypotonia — **HP:0001252**</li><li>Feeding difficulties in fancy — **HP:0008872**</li></ul> | 327 |
| 6-month | <ul><li>Poor head control — **HP:0002421**</li><li>Global developmental delay — **HP:0001263**</li></ul> | 346 |
| 12-month | <ul><li>Lazy eye – **HP:0000646**</li><li>Inability to sit — **HP:0025336**</li><li>Sleep disturbance — **HP:0002360**</li><li>Excessive drooling — **HP:0002307**</li></ul> | 91 |
| 18 month | <ul><li>Repetitive hand flapping — **HP:0100023**</li><li>Stereotypy — **HP:0000733**</li><li>Seizure-like episodes — **HP:0001250** (Seizures; confirmed later)</li><li>Absent speech — **HP:0001344**</li><li>Delayed in walking – **HP:0031936**</li><li>Motor delay – **HP:0001270**</li></ul> | 30 |
| 2 years (negative genetic tests) | <ul><li>Seizures — **HP:0001250**</li><li>Ataxia— **HP:0001251**</li><li>Abnormal EEG — **HP: 0002353**</li><li>Sleep-awake cycle disturbance — **HP: 0006979**</li></ul> | 4 |
| 3 years (diagnosis by methylation test) | <ul><li>Happy demeanor — **HP:0040082**</li><li>Inappropriate laughter — **HP:0000748**</li><li>Secondary microcephaly — **HP:0005484**</li><li>Severe intellectual disability - **HP:0010864**</li></ul> | 2 |
| 4 years+ | <ul><li>Wide mouth — **HP:0000154**</li><li>Prognathism — **HP:0000303**</li><li>Scoliosis (monitoring) — **HP:0002650**</li></ul> | 1 |